\documentclass[twocolumn,
aps,superscriptaddress,
amsmath,amssymb,
pra
]{revtex4-2}

\usepackage{amsmath}
\usepackage{bbold}
\usepackage[svgnames]{xcolor}
\usepackage{algorithm}
\usepackage{algpseudocode}
\usepackage{color}
\usepackage{amsthm}
\usepackage{graphicx}
\usepackage{dcolumn}
\usepackage{bm}
\usepackage{tikz}
\usepackage[colorlinks,breaklinks]{hyperref}
\hypersetup{
	bookmarksnumbered,
	pdfstartview={FitH},
	citecolor={darkgreen},
	linkcolor={darkred},
	urlcolor={darkblue},
	pdfpagemode={UseOutlines}}
\definecolor{darkgreen}{RGB}{50,190,50}
\definecolor{darkblue}{RGB}{0,0,190}
\definecolor{darkred}{RGB}{238,0,0}

\newtheorem{lemma}{Lemma}[section]

\newtheorem{theorem}{Theorem}[section]
\newtheorem{proposition}{Proposition}[section]
\newtheorem{definition}{Definition}[section]

\def\begeq{\begin{equation}}
\def\endeq{\end{equation}}
\def\begp{\begin{proposition}}
\def\endp{\end{proposition}}
\def\begl{\begin{lemma}}
\def\endl{\end{lemma}}
\def\begt{\begin{theorem}}
\def\endt{\end{theorem}}
\def\begd{\begin{definition}}
\def\endd{\end{definition}}

\DeclareMathOperator*{\argmax}{arg\,max}

\newcommand{\Tr}{\text{Tr}}

\newcommand{\eps}{\varepsilon}

\newcommand{\ket}[1]{\left\vert#1\right\rangle}
\newcommand{\bra}[1]{\left\langle #1 \right\vert}
\newcommand{\ketbra}[2]{| #1\rangle \langle #2|}
\newcommand{\braket}[2]{\left\langle #1 \vert #2 \right\rangle}
\newcommand{\abs}[1]{\left\vert #1 \right\vert}

\begin{document}

\title{{Choice of mutually unbiased bases and outcome labelling affects measurement outcome secrecy}}
\author{Mirdit Doda}
\affiliation{Institute of Physics, Slovak Academy of Sciences, 845 11 Bratislava, Slovakia}%
\affiliation{Institute for Quantum Optics and Quantum Information - IQOQI Vienna, Austrian Academy of Sciences, Boltzmanngasse 3, 1090 Vienna, Austria}%
\author{Matej Pivoluska}
\affiliation{Institute of Physics, Slovak Academy of Sciences, 845 11 Bratislava, Slovakia}%
\affiliation{Institute of Computer Science, Masaryk University, 602 00 Brno, Czech Republic}%
\author{Martin Plesch}
\affiliation{Institute of Physics, Slovak Academy of Sciences, 845 11 Bratislava, Slovakia}%
\affiliation{Institute of Computer Science, Masaryk University, 602 00 Brno, Czech Republic}%
\date{\today}
\begin{abstract}
Mutually unbiased bases (MUBs) are a crucial ingredient for many protocols in quantum information processing. Measurements performed in these bases are unbiased to the maximally possible extent, which is used to prove randomness or secrecy of measurement results. In this work we show that certain properties of sets of MUBs crucially depend on their specific choice, {including, somewhat surprisingly, measurement outcome labelling.} If measurements are chosen in a coherent way, the secrecy of the result can be completely lost for specific sets of MUB measurements, while partially retained for others. {This can potentially impact a broad spectrum of applications, where MUBs are utilized.}
\end{abstract}
\maketitle

\section{Introduction} 
One of the defining features of quantum mechanics is the impossibility to simultaneously measure a certain set of physical quantities. This fact led to the definition of the famous Heisenberg uncertainty principle \cite{1927ZPhy...43..172H} or understanding of the quantum  model of hydrogen atom \cite{Bohr}. 
If a simultaneous measurement of two quantities is not possible, or, in other words, if a measurement of one quantity influences the expectation of the other measurement, we call these two measurements incompatible. In this context a very natural question arises -- how much incompatible a pair of measurements can be? The answer to this question is simple --  for any quantum system,  one can find a pair of measurements where irrespective of the starting state of the system, after performing one of the measurements the result of the other one is completely random. 

A straightforward generalization is at hand -- can one form a larger set of measurements that are pairwise fully incompatible? Here again one can answer affirmatively -- for each system one can find at least three such measurements and the size of this set depends on the dimension of the system.

In order to tackle with these questions more formally, the 
notion of \emph{mutually unbiased bases} (MUBs) \cite{Schwinger570,Ivonovic_1981,PhysRevD.35.3070,WOOTTERS1989363} was introduced.  
Two $d$-dimensional bases $\{\ket{\psi_i}\}_{i = 0,\dots,d-1}$ and $\{\ket{\varphi_j}\}_{j = 0,\dots, d-1}$ corresponding to two full projective measurements are mutually unbiased, when
\begin{equation}\label{eq:MUBdefinition}
\forall i,j: \abs{\braket{\psi_i}{\varphi_j}} = \frac{1}{\sqrt{d}}.
\end{equation}
Due to their properties, 
mutually unbiased bases have become an important cornerstone of contemporary quantum information processing \cite{doi:10.1142/S0219749910006502}.
They are being used for quantum tomography \cite{Ivonovic_1981,WOOTTERS1989363}, uncertainty relations \cite{PhysRevD.35.3070,PhysRevLett.60.1103,PhysRevA.75.022319}, quantum key distribution \cite{PhysRevLett.88.127902,PhysRevA.67.062310,PhysRevA.59.4238,PhysRevA.82.030301}, quantum error correction \cite{PhysRevLett.78.405}, as well as for witnessing entanglement \cite{PhysRevA.82.012335,PhysRevA.86.022311,PhysRevLett.114.130401,PhysRevA.94.012303,PhysRevA.88.052110,Erker2017quantifyinghigh,Bavaresco2018}, design of Bell inequalities \cite{Kaniewski2019maximalnonlocality,2019arXiv191203225T} and more general forms of quantum correlations \cite{PhysRevA.92.022354,PhysRevA.95.042315,PhysRevA.98.050104}.

The natural question of the number of unbiased bases in a given dimension $d$ turned out to be unexpectedly complicated.  
While the answer is rather simple for qubits -- there are three pairwise mutually unbiased bases, defined as eigenvectors of Pauli $\sigma_x,\sigma_y,\sigma_z$ operators up to unitary equivalencies, in general, the construction of MUBs is a very difficult task.
It is known that the number of MUBs has to be smaller than $d+1$ for any dimension and the constructions of $d+1$ MUBs are known for $d=p^r$, where $p$ is a prime.
However, for non-prime-power $d$ only the trivial tensor product construction is known.

Fortunately, in many applications one needs to use only $k\leq d+1$ MUB measurements. Clearly, there are different ways to pick the subset of $k$ out of all MUBs. In fact, it is known that different sets of MUBs are not necessarily equivalent under different mathematical operations, such as global unitary operations, changing individual vector phases, relabelling of outcomes, relabelling of moments or introducing complex conjugation \cite{DBLP:journals/qic/BrierleyWB10}. This mathematical inequivalence is however irrelevant in many practical applications where just satisfying the defining property \eqref{eq:MUBdefinition} is required for the task.

More interestingly, it was recently shown that different subsets of MUBs of can be inequivalent operationally as well. 
For example, MUBs turn out to be an optimal strategy in a communication task called quantum random access coding (QRAC) \cite{PhysRevA.99.032316}. 

In \cite{PhysRevLett.121.050501} it was shown that in a certain variant of QRAC, different subsets of $k$ out of $d+1$ MUBs lead to different strategies with different average success rates. 
More recently, it was shown that different subsets of $k$ out of $d+1$ MUBs behave differently under a measure called incompatibility robustness \cite{PhysRevLett.122.050402}. Last but not least, very specific MUBs are required to obtain Bell inequalities \cite{Kaniewski2019maximalnonlocality}, which are maximally violated by maximally entangled states and MUBs.

The full definition of a measurement consist of specifying the basis as a set of states and labelling these states. Two measurements consisting of the same set of states are in principle different, even if they measure the same property and their results can be classically transformed at any later stage.  From the experimental and operational point of view it makes sense to distinguish between different measurements that only differ in labelling (we call this a \emph{classical difference}) and two measurements that differ in the states per se (\emph{quantum difference}). 
One can then naturally ask, to what extent the properties of MUBs do change if one only makes a classical change in them. 
In other words, do the properties of the subsets change by simple re-labelling of their vectors?
In this work, we affirmatively answer this question by introducing a quantum information task called \emph{guessing game}. 
There a subset of $d$ out of $d+1$ MUBs is used to hide and guess information between two parties. 
{We show that this simple choice of removing a single MUB from the full set critically affects achievable results in the game. }
Even more interestingly, for a suitable chosen subset of $d$ out of $d+1$ MUBs,  {we observe} the full spectrum of results --  perfect guessing and maximal hiding -- just by relabelling the measurement outcomes. 


\section{Results} 
The incompatibility of measurements can be demonstrated and examined with the help of a very simple quantum game, studied in \cite{Rozp_dek_2017,Plesch_2018}. Here Alice realizes one of $m$ possible measurements on a $d$-dimensional system and records the result $a$ of this measurement. The task of Bob is to guess this result using the following strategy: first, he prepares the state for Alice to be measured and second, he receives information about which measurement was performed (see the next section for the full definition of the guessing game). 

If the game is described by classical physics, a pure state has a determined outcome for all possible measurements. 
Therefore, trivially, Bob can prepare a state which leads to a deterministic outcome irrespective on measurement performed by Alice.

One can make the scenario partially quantum, by making Bob's probe state as well as the measurements quantum, but keep the information about the measurement chosen by Alice classical -- we call this a \textit{classical coin scenario}.
This is the traditional way to demonstrate incompatibility of quantum measurements -- for compatible measurements Bob still can guess with certainly, but with increasing incompatibility of the measurements the uncertainty of his guess increases.

\begin{figure} 
\centering
\begin{tikzpicture}[scale = 1.45]
\node at (-2,2.5) {\LARGE{\bf$\rho_B$}};
\node at (-1,1) {\LARGE{\bf{$\rho_C$}}};
\draw (-1.5,2.5) -- (0.5,2.5);
\draw  [thick](0.5,3) rectangle (1.5,2);
\node [thick] at (1,2.5) {\LARGE{$U^\dagger_i$}};
\draw  [thick,fill](1,1) ellipse (0.1 and 0.1);
\draw (-0.5,1) -- (1.1,1);
\draw (1,1) -- (2.5,1);
\draw (1.5,2.5) -- (2.5,2.5);
\draw (1,2) -- (1,1);
\draw (2.5,1.5) arc (90:-90:0.5);
\draw (2.5,3) arc (90:-90:0.5);
\draw (2.5,3) -- (2.5,2);
\draw (2.5,1.5) -- (2.5,0.5);
\node at (3.25,2.5) {\LARGE$a$};
\node at (3.25,1) {\LARGE$b$};
\node at (2.75,1) {\large{$M_b$}};
\end{tikzpicture}
\caption{Guessing game description. Alice measures the probe state $\rho_B$ with one out of $d$ possible measurements. Alice's measurements choice is implemented coherently, via a controlled unitary $\sum_{i=0}^{d-1}U_i^{\dagger}\otimes  \ketbra{i}{i}$, where $U_i^{\dagger}$ maps the basis vectors of the $i$-th basis onto the computational basis. Alice then measures in the computational basis and her outcome is denoted $a$. Bob's goal is to guess Alice's outcome by preparing a probe state $\rho_B$ and an optimal measurement described by POVM elements $\{M_b\}_{b=0}^{d-1}$, through which he obtains his guess $b$. Bob wins when $b=a$. In the \textit{classical coin} case, the control state $\rho_C$ is fully mixed , and in the \textit{quantum coin} case, $\rho_C$ is a superposition of computational basis vectors.}
\label{fig:setup}
\end{figure}
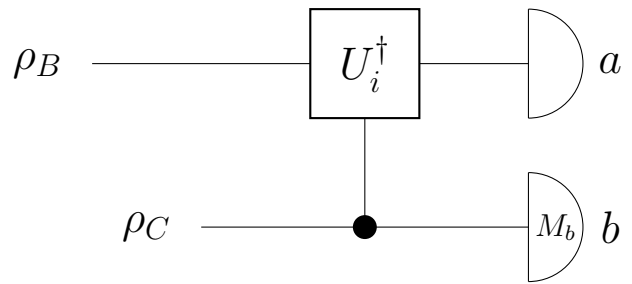

In a fully quantum scenario -- called \textit{quantum coin scenario} -- depicted in Figure (\ref{fig:setup}),
both the probe state and the information about the measurement chosen are quantum. Here Alice realizes the chosen measurement by first applying a coherently controlled
unitary, followed by a measurement in a standard basis. Bob receives the control state and can use it to determine Alice's outcome. 

The authors of \cite{Rozp_dek_2017} have analyzed the guessing game for two specific MUB measurements ($m=2$). 
They have shown that for qubits ($d=2$), in the quantum coin scenario Bob can guess Alice's outcome with certainty.
In contrast, this was not the case for higher dimensions. 
 {They have concluded that} in case of two measurements the control state is always a two dimensional state  and it is impossible to use it to determine a higher dimensional outcome.

In \cite{Plesch_2018} we have further analyzed the guessing game with the quantum coin and we have shown that for qubits, with any number of measurements (independent on their level of compatibility) it is always possible for Bob to obtain the result of Alice with probability $1$. 
In contrast, for higher dimensions this is not the case, so even if Bob receives a large enough control state, he will not be able to guess the result perfectly for a \emph{specific set of MUBs} chosen by Alice.

Here we analyze the problem further. We fix the number of measurements to $m=d$, which will make the size of the measurement outcomes alphabet equal to the dimension of the control state available to Bob.
{First we study quantum coin scenario with this choice for different sets of $d$ MUBs and for each prime $d$ we construct a set of $d$ MUBs, which allow Bob to guess Alice's measurement outcomes with certainty. 
Further, with a combination of exhaustive search for $d=3$ and $d=5$ and numerical methods for higher dimensions we study Bob's guessing probability with different sets of $d$ MUB measurements. We consider MUBs obtained by choosing $d$ out of $d+1$ MUBs from standard Wootters-Fields (WF) construction (see \cite{WOOTTERS1989363} and equation \eqref{WF}) followed by relabelling of their vectors in order to obtain different measurements.}  

{Strikingly, both the lowest and the highest guessing probabilities we observe are achieved by excluding the computational basis from $d+1$ WF bases and imposing different labelling of measurement outcomes to the rest of bases -- original WF labelling leads to  the lowest guessing probabilities while our construction, which is yet another outcome relabelling of this set of MUBs, leads to perfect guessing probability.
More broadly, our study goes far beyond the study of the guessing game itself, as it shows that different sets of $d$ out of $d+1$ MUBs, which only differ in a classical sense (i.e. by relabelling), exhibit very different operational properties. }

\section{Guessing game} 
Here we give a formal definition of the guessing game and define a set of $d$ out of $d+1$ MUB measurements which allows Bob to construct a perfect guessing strategy. 
In the guessing game, Alice receives an initial state $\rho_B$ of dimension $d$ prepared by Bob. 
She performs a coherently controlled unitary transformation $CU$ defined by the set of $\{U_a^{\dag}\}_{a=0}^{d-1}$ controlled by the ``coin'' state $\rho_C$. 
In the quantum coin scenario the pure state $\rho_C = \ketbra{+}{+}$ is used, where $\ket{+} = \tfrac{1}{\sqrt{d}}\sum_{i=0}^{d-1} \ket{i}$, while in the classical coin scenario a fully mixed state $\rho_C=\frac{\mathbb{1}}{d}$ is used. 
After the transformation, Alice measures the state ${\rho_B}$ in the computational basis and sends the control state $\rho_C$ to Bob, who also performs a general measurement defined by POVM elements $\{M_b\}_{b=0}^{d-1}$ to obtain his guess $b$. 
Bob wins if the results coincide. 

The average guessing probability of Bob is defined as:
\begin{equation}
    P_g\!\!:=\!\!\sum_{a=0}^{d-1}  \Tr\left[   (\rho_B \!\otimes\! \rho_C) CU \left(\ketbra{a}{a} \otimes M_a\right)CU^{\dag}\right],
\end{equation}

Although there are multiple constructions of MUBs for prime dimensions, to demonstrate our result we will use a construction of Wootters and Fields (WF) \cite{WOOTTERS1989363}: 
\begin{equation} \label{WF}
U^{\text{WF}}_a=\frac{1}{\sqrt{d}}\sum_{i,j=0}^{d-1} \omega^{ai^2+ij}\ketbra{i}{j}.
\end{equation}
In prime dimension $d$, this construction defines $d$ different bases and can be supplemented by the computational basis to for the full set of $d+1$ MUBs. There are $d+1$ different ways to select the set of $d$ bases. Additionally, for each set of $d$ bases we will consider relabelling of the vectors which allows us to construct additional sets of $d$ measurements used in the guessing game. 

\subsection{Classical coin scenario} 
In the case of a classical coin state, we have that $\rho_C=\frac{\mathbb{1}}{d}$. Clearly, this is equivalent to Alice choosing the measurement uniformly at random and Bob then receiving the information about which measurement was chosen. Based on this information he has to guess the result obtained by Alice. While for qubits the optimal strategy for Bob is straightforward and easy to understand (he prepares a coherent superposition of two basis states of the two possible measurements of Alice) and yields the guessing probability of $\frac{1}{2}\left( 1+\frac{1}{\sqrt{2}} \right)$, for the higher dimensional variant of the game the situation is much more complicated. 
In Appendix \ref{app:classical} we derive an \emph{upper bound} in the form $\frac{1}{d}\left( 1+\frac{d-1}{\sqrt{d}} \right) $ valid for \textit{any} set of MUBs (this includes relabelling, since it does not influence the Bob's guessing probability in the classical coin scenario), which converges to $0$ for high $d$.
Furthermore, for the set of MUBs  defined in (\ref{WF}) up to $d=7$ we also obtain exact values. 
For higher $d$ we provide numerical estimates that show that the bound obtained is not tight.  {These results show} that without coherent information, with increasing $d$, Bob can only obtain negligible information about the result obtained by Alice irrespective on which set of MUBs she uses. 


\subsection{Quantum coin scenario} 
The situation is dramatically different for the quantum coin scenario, in which $\rho_C = \ketbra{+}{+}$. First we show that for a specific selection of MUBs  it is possible for Bob to obtain  {Alice's result} with certainty. To achieve this, Alice needs to select both the proper $d$ WF MUBs, (quantum setting) and label the individual measurement basis vectors in a suitable way as well (classical setting). Specifically, if Alice chooses $d$ WF bases without relabelling, Bob can never achieve perfect guessing, as we have shown in \cite{Plesch_2018}.

{MUBs which result in Bob's perfect guessing probability are defined as}  
\begin{align} \label{perfectunitaries}
     U_a^{DPP}&=\frac{1}{\sqrt{d}}\sum_{i,j=0}^{d-1}\omega^{a i^2 + ij - a^2 i}\ketbra{i}{j},
\end{align}
which {can be seen as} relabelling of the vectors of the bases of the WF construction: $U_a^{DPP}\ket{j}=U^\text{WF}_a\ket{j-a^2}$.

Let us define Bob's (pure) probe state $\ket{\psi_B}$ and measurements $\{M_a\}_{a=0}^{d-1}$ as:
\begin{align} \label{optimalconstruction}
\begin{split}
        \ket{\psi_B}&=\frac{1}{\sqrt{d}}\sum_{k=0}^{d-1} \omega^{ 3^{d-2} k^3} \ket{k},\\
M_a&=\frac{\ketbra{\phi_a}{\phi_a}}{\braket{\phi_a}{\phi_a}}, \\
  \ket{\phi_a}&:=\frac{1}{\sqrt{d}}\sum_{a=0}^{d-1} \bra{j} U_a^{\dag} \ket{\psi_B} \ket{a},
  \end{split}
\end{align}
where $\ket{\phi_j}$ are unnormalized pure states.
In the Appendix \ref{app:proof} we show that $\{M_a\}_{a=0}^{d-1}$ form a projective measurement. Subsequently we show that such measurement allows Bob to guess perfectly Alice's measurement outcomes if used in conjunction with the probe state $\ket{\psi_B}$.

Interestingly, if one of the WF bases is exchanged for the computational basis (which corresponds to a quantum difference), there is no way for Bob to achieve perfect guessing for any labelling of the individual measurements. In other words, if the computational basis is included in the set of MUBs used, we have strong numerical evidence that Alice can retain some secrecy towards Bob irrespective of the labelling used; for dimensions $3$ and $5$ this can be shown by exhaustive search over all the possible relabellings, for higher dimensions we performed a randomized search (see Appendix \ref{app:numSearch} for details). 

\section{Optimal hiding in the quantum coin case} 
We have shown that if Bob can influence the choice of MUBs used by Alice, he can perfectly guess her outcome. It is thus very natural to ask the complementary question -- if Alice can retain full control about her measurements, what is the maximum Bob can learn about her outcome? And how does this maximum depend on the quantum setting of her measurements and actual labelling?

To answer this question fully, one would have to search through all possible MUBs including their labelling and find optimal values. To keep the task tractable, first we have focused on the standard WF set of MUBs plus the computation basis (leading to $d+1$ possibilities) plus possible relabellings expressed via permutation matrices $P_\pi$, which relabel the computational basis states and leave the MUB property intact:
\begin{align*}
     \abs{\bra{i} U_a^{\dag} U_b \ket{j}}& = \frac{1}{\sqrt{d}}=  \abs{\bra{i} P_{\pi} U_a^{\dag} U_b P_{\pi '} \ket{j}}.
\end{align*}
Due to the intractably large number of combinations, for dimensions higher than $5$ we first restricted ourselves to cyclic permutation matrices. On top of it, we have also tested randomly a large set of non-cyclic permutation matrices.

\begin{figure}[ht]
     \centering
     \includegraphics[width= 0.47\textwidth]{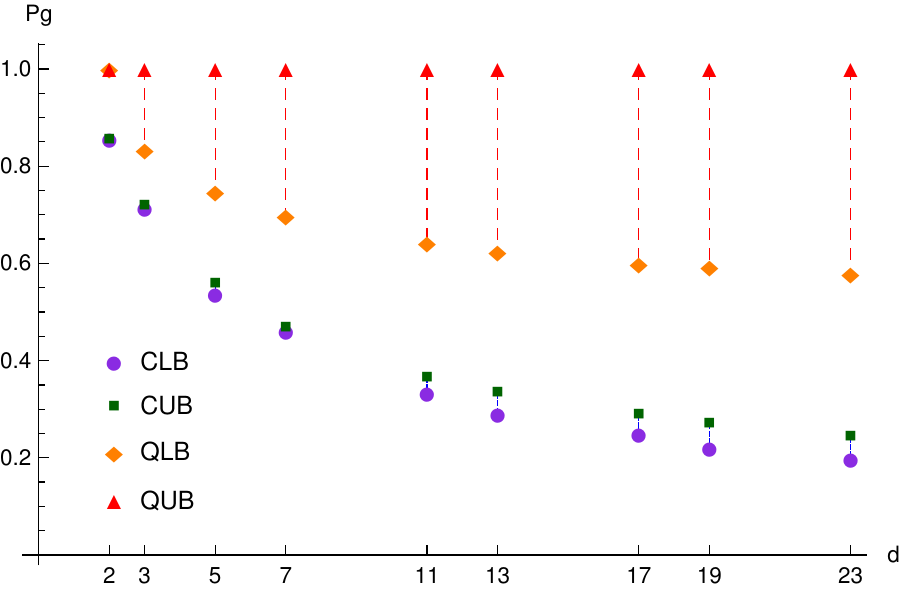}
     \caption{
   Here we depict the bounds of the guessing probability for the classical and quantum coin for different dimensions. The quantum coin upper bound (QUB) is analytical and equal to $1$. For $d$ up to $5$ the quantum coin lower bounds (QLB) are over all relabelling (permutations), for higher dimensions over all cyclic permutation topped up by a random search. For $d$ up to $7$ the classical coin lower bounds (CLB) are tight and obtained by an exhaustive search. The classical upper bounds (CUB) are obtained via matrix inequalities. }
     \label{plot}
 \end{figure}

For a fixed set of MUBs, we cast the problem as a see-saw SDP \cite{LAURENT2005393}(see Appendix \ref{app:SSSDP} for details), which allows us to obtain a lower-bound on  $P_g$. We have randomized the initial point and repeated the optimization to obtain the lower bounds as depicted in Figure (\ref{plot}).
As a last step, to look a bit behind the strict limit of WF construction and its relabelling, we applied the see-saw algorithm to unitaries close to the MUBs in the space of unitary matrices. In all cases we obtained values higher than the WF construction; this shows that the found values constitute (at least) a local minimum in the space of unitary matrices, while the search over permutation matrices suggests that they constitute a global minimum over the space of MUB unitary matrices as well.


While the obtained minima decrease with the dimension, they stay far above the upper bounds of the classical coin scenario. Thus it is clear that irrespective of the selection of measurements by Alice, obtaining coherent information about her measurement allows Bob to take a more accurate guess.
{At the same time, in the case of the quantum coin, the maximal and minimal guessing probability discovered with our numerical methods change with the choice of both measurement bases and their labelling, making it critically important for Alice to carefully choose the MUBs used in the guessing game.} 

An analysis of the actual MUBs that lead to the obtained minimum guessing probability sheds some light on the problem. Surprisingly, it turned out that the minimal guessing probabilities {we found} are obtained for the standard WF construction of MUBs $\{U^{\text{WF}}_a\}_{a=0}^{d-1}$. So in the case when Alice can make her choice of the measurements, including the labelling, it is best for her to select the standard construction to minimize the knowledge of Bob. At the same time we could see that the perfect guessing by Bob was achieved for the DPP construction (\ref{perfectunitaries}), which only differs from the WF construction by relabelling -- i.e. boundary values {we found} are achieved for MUBs that differ only by labelling. 
 
 On the contrary, if the computational basis is included into the system by exchanging it with any of the WF bases, we have strong numerical evidence that neither Bob can perfectly guess the outcome, nor Alice can hide it as well as in the WF case. This suggests that the set of $d$ WF constructed bases including its relabelling is structurally different than any set where the computational basis is used with $d-1$ WF bases. It is worth mentioning that this fact is not connected to the computation basis itself. One can find sets of MUBs containing computation basis that exhibit the same properties as the WF or DPP set respectively, but the remaining bases are not given by the WF construction. 

\section{Discussion} 
In our work we have shown, using a simple quantum mechanical game, that different choices of mutually unbiased bases have dramatic effects on experimentally achievable results. Interestingly, for any prime dimension $d$ one can choose a set of $d$ MUBs that provide the possibility of perfect guessing by Bob of the result obtained by Alice in the quantum coin scenario. At the same time, {we obtained a strong numerical evidence that} with a set of MUBs that differs only by relabelling of the individual vectors, Alice can obtain the maximum hiding of her result the game allows. 

This result is very striking on its own, as it shows a very interesting and deep structure of the seemingly simple construction of MUBs. Even though all of the bases look very similar in its mathematical form, the subtle phase interdependences allow for some of the subsets to deliver truly different results than others. 

More than that, the result is interesting from a practical viewpoint as well. While it might be considered as very artificial to introduce a quantum control of the measurement chosen by Alice, this is in fact the way how such a control works for instance on the IBM quantum computer, where no classical control is available \footnote{An example of using coherent control instead of classical is given on qiskit tutorial on quantum teleportation. URL: \url{https://qiskit.org/textbook/ch-algorithms/teleportation.html#4.1-IBM-hardware-and-Deferred-Measurement-}}. In the future design of quantum security elements it is  possible that due to technological reasons, quantum controls will be a standard procedure. In such a case, it will be very important to carefully consider the design of the quantum part so that the selected MUBs are not only secure as designed, but are (reasonably) secure even in the case of coherent control and possible relabelling. 

\begin{acknowledgments}
\textit{Acknowledgements.} We would like to thank Flavien Hirsch and Marco T\'{u}lio Quintino for innitial discussions and M\'{a}t\'{e} Farkas and Jed Kaniewski for discussions about MUBs. We acknowledge funding from VEGA project 2/0136/19. MPi and MPl additionally acknowledge GAMU project MUNI/G/1596/2019.    
\end{acknowledgments}

\appendix
\section{Optimization Algorithm}
\label{app:SSSDP}
Given a MUB construction encoded by the unitaries $\{U_a\}_{a=0}^{d-1}$, we want to estimate the associated optimal strategy that Bob can use to guess Alice's outcomes in the quantum coin scenario. The optimal strategy would be the result of the following optimization:
\begin{align}
\begin{split}
    \label{nloptimization}
   P_g^{\text{max}}\!\!&=\!\!\!\!\!\!\max_{\rho_B,\{M_a\}_{a=0}^{d-1}}
   \sum_{a=0}^{d-1} \! \Tr_{AB}\left[  \! (\rho_B \!\otimes\! \rho_C)CU\!\! \left(\ketbra{a}{a}\! \otimes\! M_a\right)\!CU^{\dag}\right]\\
    &\text{s.t.} \hspace{0.3cm}  \rho_B \geq 0\\
    &\Tr \rho_B =1\\
    &M_a \geq0 \hspace{0.3cm}\forall a\in \{0,\ldots,d-1\}\\
    &\sum_{a=0}^{d-1}M_a=\mathbb{1},
    \end{split}
\end{align}
where the optimization variables are Bob's probe state $\rho_B$ and Bob's POVM elements $M_a$ corresponding to the outcome $a$. Also recall that $\rho_C$ is the control state representing the choice of measurements, and $CU$ is a controlled unitary used to implement Alice's measurement settings coherently.
The target function of this optimization problem is non-linear, therefore it cannot be solved directly by Semi-Definite Programming (SDP). We therefore cast it as two SDPs, which we run alternatively. In the first SDP
we optimize over $\{M_a\}_{a=0}^{d-1}$ with $\rho_B$ constant and in the second one we optimise over $\rho_B$ while $\{M_a\}_{a=0}^{d-1}$ are constant:

\begin{align*}
   \text{{\bf SDP 1: }}&\text{given }  \rho_B
\\
   \{ M_a\}_{a=0}^{d-1}=         & \argmax_{\{M_a\}_{a=0}^{d-1}} 
   \frac{1}{d}\sum_{i,j,a=0}^{d-1} \bra{i}M_a\ket{j} \bra{a} U_j^{\dag} \rho_B U_i \ket{a}
\\
    \text{s.t.} \hspace{0.3cm}   & M_a \geq0 \hspace{0.3cm}\forall a\in \{0,\ldots,d-1\}
\\
                                 & \sum_{a=0}^{d-1}M_a=\mathbb{1}
\end{align*}
\begin{align*}
   \text{{\bf SDP 2: }}&\text{given }     \{ M_a\}_{a=0}^{d-1}\\
    \rho_B=                       & \argmax_{\rho_B}
   \frac{1}{d}\sum_{i,j,a=0}^{d-1} \bra{i}M_a\ket{j} \bra{a} U_j^{\dag} \rho_B U_i \ket{a}\\
   \text{s.t.} \hspace{0.3cm}   & \rho_B \geq 0\\
        & \Tr \rho_B =1,
\end{align*}
where we simplified the notation with 
\begin{align*}
 \sum_{a=0}^{d-1}  &\Tr_{AB}\left[   (\rho_B \!\otimes\! \rho_C) CU \left(\ketbra{a}{a} \otimes M_a\right)CU^{\dag}\right]
 =\\
 &\frac{1}{d}\sum_{i,j,a=0}^{d-1} \bra{i}M_a\ket{j} \bra{a} U_j^{\dag} \rho_B U_i \ket{a},\\
 \end{align*}
 for
 \begin{align*}
      CU&=\sum_{i=0}^{d-1}U_i^\dagger\otimes  \ketbra{i}{i} 
 ,\\
 \rho_C&=\ketbra{+}{+}
 ,\\
 \ket{+}&=\frac{1}{\sqrt{d}}\sum_{i=0}^{d-1} \ket{i}. 
 \end{align*}

The two SDPs are each guaranteed to converge, the see-saw, however must stop at a `convergence parameter' $\eps$ that we set to be $10^{-6}$; explicitly, the see-saw algorithm is the following:

\begin{algorithm}[H]\label{algorithm}
	\floatname{algorithm}{Algorithm}
	\caption{See-saw}
	\begin{algorithmic}[1]
\State\label{generate}\noindent\textbf{Initialization:} 
Generate a random density matrix $\rho_0$, distributed according to the Hilbert-Schmidt measure. Set $P_W=0$.
\State\label{SDP1}\noindent\textbf{POVM optimization:} 
Given $\rho_0$, solve the SDP with $\{M_a\}_{a=0}^{d-1}$ as variable, and find the solution $\{M_a^*\}_{a=0}^{d-1}$.
\State\label{SDP2}\noindent\textbf{State optimization:}
Given $\{M_a^*\}_{a=0}^{d-1}$ from step 2, solve the SDP with $\rho_B$ as variable, and find the solution $\rho_B^*$ and $P_W^*$.
\State\label{conv}\noindent\textbf{Convergence check:} 
\begin{itemize}
\item If $P_W^*-P_W>\eps$, then set $\rho_0=\rho_B^*$ and $P_W=P_W^*$. Repeat from step 2.
\item If $P_W^*-P_W<\eps$, stop the algorithm. The complete solution is given by $P_W^*$, $\rho_B^*$, $\{M_a^*\}_{a=0}^{d-1}$.
\end{itemize}
	\end{algorithmic}
\end{algorithm}
The algorithm is then applied to a large number of initial random points $\rho_0$. We observed that for $\eps$ small enough it yields always the same result $P_W^*$, suggesting that the see-saw algorithm lower bounds tightly the solution of (\ref{nloptimization}).

\section{Optimal strategy}
\label{app:proof}

In the DDP construction we considered Alice's MUB measurements defined as $U_a          =\frac{1}{\sqrt{d}} \sum_{i,j=0}^{d-1} \omega^{a i^2 + ij - a^2i } \ketbra{i}{j}$. 
Bob's optimal strategy in this case is:
\begin{align*}
\ket{\psi_B} &=\frac{1}{\sqrt{d}} \sum_{k=0}^{d-1} \omega^{ 3^{d-2} k^3} \ket{k},\\
    M_a          &=\frac{\ketbra{\phi_a}{\phi_a}}{\braket{\phi_a}{\phi_a}},\\
\ket{\phi_a} &=\frac{1}{\sqrt{d}}\sum_{a=0}^{d-1} \bra{j} U_a^{\dag}   \ket{\psi_B} \ket{a},\\
\end{align*}
where $\ket{\psi_B}$ is Bob's (pure) probe state and $\{M_a\}_{a=0}^{d-1}$ are POVM elements of the measurement he uses on the probe state $\rho_C = \ketbra{+}{+}$ to guess Alice's outcome. Note that states $\ket{\phi_a}$ are not normalized.

Here we show that $\{M_a\}_{a=0}^{d-1}$ is indeed a valid POVM, i.e. $M_a\geq0 \hspace{0.2cm}\forall a$ and $ \sum_{a=0}^{d-1}M_a=\mathbb{1}$. 
Positivity is guaranteed by definition. To prove summation to identity we notice that $M_a$ are projectors and span the Hilbert space of Bob if $\left\{ \frac{\ket{\phi_j}}{ \| \ket{\phi_j}\|} \right\}_{j=0}^{d-1}$ form an orthonormal basis. Normalization is guaranteed by definition, so it remains to prove orthogonality:
\begin{align*}
    &\braket{\phi_i}{\phi_j}
    =\frac{1}{d}\sum_{a=0}^{d-1}  \bra{j} U_a^{\dag}   \ket{\psi_B}  \bra{\psi_B} U_a   \ket{i}=\\&
    =\frac{1}{d^3}\sum_{a,k,l=0}^{d-1} \omega^{-(a k^2 + jk - a^2 k)}  \omega^{ 3^{d-2} k^3} \omega^{- 3^{d-2}l^3} \omega^{a l^2 + il - a^2 l}\\
    &=\frac{1}{d^3}\sum_{a,k,l=0}^{d-1} \omega^{-a k^2 - jk + a^2 k+ 3^{d-2} k^3- 3^{d-2}l^3+a l^2 + il - a^2 l}.
\end{align*}
\subsection{Dimensions larger than 3}
In what follows, we will show that for $d>3$ ($d=3$ and $d=2$ are treated separately) the above expression can be simplified using quadratic Gauss sums. In order to do so, we will manipulate the exponents of $\omega$. The key idea is to realize that since $\omega^d = 1$ , we can work with its exponent modulo $d$. Additionally, we introduce a substitution:
\begin{align*}
    m=l+k \quad \text{and}\quad n=l-k, 
\end{align*}
and two constants 
\begin{align*}
    \alpha &= 3^{d-2} \equiv 3^{-1}\pmod{d}\quad\\\quad \beta 
    &= 2^{d-2} \equiv 2^{-1} \pmod{d}.
\end{align*}
From these definitions it follows that 
\begin{align*}
   l&\equiv \beta(m+n)\pmod{d}                 ,\\     3\alpha &\equiv 1\pmod{d},\\
   k&\equiv \beta(m-n) \pmod{d}                ,\\    2\beta &\equiv 1\pmod{d},\\
   l^2-k^2&\equiv mn\pmod{d}                   ,\\    il-jk&\equiv \beta m(i-j)+\beta n(i+j)\pmod{d} ,\\
   l^3-k^3&\equiv \beta^2n(3m^2+n^2)\pmod{d}  .&   & 
\end{align*}
We will also use the quadratic Gauss sum:
\begin{align*}
    \sum_{a=0}^{d-1} \omega^{a^2 m}&=
    \begin{cases}
    \left(\frac{m}{d}\right) \varepsilon_d \sqrt{d} \hspace{0.5cm} &\text{if }m \not\equiv 0  \text{ (mod } d)\\
    d \hspace{0.5cm} &\text{if }m \equiv 0  \text{ (mod } d)
    \end{cases},
\end{align*}
where $\left(\frac{m}{d}\right)$ is the Legendre symbol:
\begin{align*}
    \left(\frac{m}{d}\right)&
    =\begin{cases}
    1 \hspace{0.5cm} &\text{if }\exists n:   m \equiv n^2  \text{ (mod } d)\\
    -1 \hspace{0.5cm} &\text{if }\nexists n:  m \equiv n^2  \text{ (mod } d)
    \end{cases} ,
\end{align*}
and
\begin{align*}
     \varepsilon_d =
     \begin{cases} 
     1 \hspace{0.5cm}  &\text{if $d \equiv 1$   (mod $4$)}\\
     i\hspace{0.5cm}  &\text{if $d \equiv 3$   (mod $4$)}
     \end{cases}.
\end{align*}
 After the substitution, the expression reads:
\begin{align*}
     &\braket{\phi_i}{\phi_j} = \\
     &=\!\!\frac{1}{d^3}\!\!\sum_{a,m,n=0}^{d-1}\!\!\!\!\omega^{amn-a^2n-\alpha \beta^2 n^3 -\beta^2  m^2 n +\beta m (i-j) + \beta n (i+j)}\\
     &=\!\!\frac{1}{d^3}\!\!\sum_{m,n=0}^{d-1}\!\!\!\!\omega^{-\alpha \beta^2 n^3 -\beta^2  m^2 n +\beta m (i-j) + \beta n (i+j)}\sum_{a = 0}^{d-1} \omega^{amn-a^2n}.
\end{align*}
The sum over $a$ is a quadratic Gauss sum:
\begin{align*}
    \sum_{a=0}^{d-1} &\omega^{-a^2n+amn}
    \\
    &=    \sum_{a=0}^{d-1} \omega^{-n(a-\beta m)^2} \omega^{  \beta^2 m^2 n}\\
    &=  \omega^{  \beta^2 m^2 n}\sum_{a=0}^{d-1} \omega^{-a^2 n}\\
    &=     
    \begin{cases}
    \omega^{ \beta^2  m^2 n }\left(\frac{-n}{d}\right)\varepsilon_d \sqrt{d}\hspace{0.5cm} &\text{if }n \not\equiv 0  \text{ (mod } d)\\
    d \hspace{0.5cm} 
    &\text{if }n \equiv 0  \text{ (mod } d)
    \end{cases},
    \end{align*}
    where the second equality follows from the fact that $(a-\beta m)^2$ iterates over the same values $\pmod d$ as $a^2$.
Substituting this expression in the previous one, we obtain:
\begin{align}
\begin{split}
\label{Proof1}
     \braket{\phi_i}{\phi_j}
     & =\frac{1}{d^{3}} \sum_{m=0}^{d-1} \omega^{\beta m (i-j)} \\&\quad\times\left[\sum_{n=1}^{d-1}\varepsilon_d \sqrt{d}  \left(\frac{-n}{d}\right)\!\omega^{ -\alpha \beta^2 n^3 + \beta n (i+j)}\!+\!d\right]=\\
        &=\frac{\delta_{ij}}{d} \left[  \frac{\varepsilon_d}{ \sqrt{d}} \sum_{n=1}^{d-1} \left(\frac{n}{d}\right) \omega^{ 12^{(d-2)} n^3 -  n j}+1\right],
        \end{split}
\end{align}
which shows that they are orthogonal as requested. We then show that this construction gives a guessing probability $P_g = 1$:
\begin{align}
\begin{split} \label{proof2}
    P_g&=
     \sum_{k=0}^{d-1}  \Tr_{AB}\left[ CU^{\dag}  (\rho_B \!\otimes\! \rho_C) CU \left(\ketbra{k}{k} \otimes M_k\right)\right]
    \\
 &
    =\sum_{k=0}^{d-1}\!\Tr_{AB}\!\!
    \left(\sum_{a=0}^{d-1} U_a^{\dag}\!\otimes\!\ketbra{a}{a}\!\!\right)
    \!\!\!\left(\!\!\ketbra{\psi_B} {\psi_B}\!\otimes\!\frac{1}{d} \sum_{i,j=0}^{d-1}\!\ketbra{i}{j}\!\!\right)\\
    &\quad\quad\times
    \left( \sum_{b=0}^{d-1} U_b  \otimes  \ketbra{b}{b}  \right)
    \left( \ketbra{k}{k} \otimes \frac{\ketbra{\phi_k}{\phi_k}}{\braket{\phi_k}{\phi_k}}\right)
    \\&
    = \Tr_{B}\sum_{k=0}^{d-1}
    \left(\frac{1}{\sqrt{d}}\sum_a \bra{k}U_a^{\dag} \ket{\psi_B} \ket{a}\right)\\
    &\quad\quad\times
    \left(\frac{1}{\sqrt{d}}\sum_b    \bra{\psi_B}U_b\ket{k} \bra{b}\right) 
    \frac{\ketbra{\phi_k}{\phi_k}} {\braket{\phi_k}{\phi_k}}
    \\&
    =  \Tr_{B}\sum_{k=0}^{d-1}
    \ketbra{\phi_k}{\phi_k}\frac{\ketbra{\phi_k}{\phi_k}}{\braket{\phi_k}{\phi_k}}
    \\&= \sum_k \braket{\phi_k}{\phi_k}
    \\&
    =\frac{1}{d}\sum_{a,b,k=0}^{d-1}\braket{b}{a}\bra{k}U_a^{\dag} \ket{\psi_B}  \bra{\psi_B}U_b\ket{k}\\
    &=\frac{1}{d}\sum_{a=0}^{d-1} \bra{\psi_B}U_a\left(\sum_{k=0}^{d-1}\ketbra{k}{k}\right)U_a^{\dag}\ket{\psi_B}
    \\&
    =\frac{1}{d}\sum_{a=0}^{d-1} \bra{\psi_B} \mathbb{1}\ket{\psi_B}
    =1.
    \end{split}
\end{align}

\subsection{Dimension 3}
\label{app:proof3}
Above we have shown that Bob can guess with probability one for $d>3$. For the case $d=2$ the optimal strategy can be found in \cite{Rozp_dek_2017}. For $d=3$, the proof needs to be adapted due to the fact that a multiplicative inverse $\pmod{3}$ of $3$ does not exist; we then use $\omega=e^{\frac{2\pi i}{3}}$ for Alice's MUB construction $U_a =\frac{1}{\sqrt{d}} \sum_{i,j=0}^{d-1} \omega^{a i^2 + ij - a^2i } \ketbra{i}{j}$ and with $\omega_9=e^{\frac{2\pi i}{9}}$ we define Bob's strategy as:
\begin{align*}
  M_a          &=\frac{\ketbra{\phi_a}{\phi_a}}{\braket{\phi_a}{\phi_a}},\\
  \ket{\phi_a} &=\frac{1}{\sqrt{d}}\sum_{a=0}^2 \bra{j} U_a^{\dag}   \ket{\psi_B} \ket{a},\\
  \ket{\psi_B} &=\frac{1}{\sqrt{d}} \sum_{k=0}^2 \omega_9^{ k^3} \ket{k},\\
\end{align*}
The proof follows exactly the same steps of \ref{Proof1}, with all the substitutions remaining valid, with the exception of
\begin{align*}
    \omega_9^{k^3-l^3}&= \omega_9^{-\beta^2 n(3m^2+ n^2)}\\&=\omega^{-\beta^2 m^2 n} \omega_9^{-\beta^2 n^3}\\&=\omega^{-\beta^2 m^2 n} \omega_9^{-7 n^3},
\end{align*}
where we made use of the fact that $\beta=5$ is the multiplicative inverse of $2 \pmod{3}$ and$\pmod{9}$. We then get
\begin{align*}
      \braket{\phi_i}{\phi_j}
     =\frac{\delta_{ij}}{3} \left[ \frac{\varepsilon_3}{ \sqrt{3}} \sum_{n=1}^{2} \left(\frac{n}{d}\right) \omega_9^{ 7 n^3 - 3 n j}+1\right],
\end{align*}
concluding the proof.

\section{Classical coin}
\label{app:classical}
In the classical case, the control state is a computational basis vector $\ket{i}$, chosen uniformly at random, which selects the measurement used by Alice via controlled unitary $CU$. Therefore, it contains full information about the basis Alice measures in, which can be obtained by Bob performing a measurement in computational basis. Any other measurement by Bob only introduces extra entropy to this information via uncertainty principle and thus decreases Bob's guessing probability.
Bob's optimal guessing strategy is therefore a simple projection onto the computational basis, which reveals Alice's measurement basis $i$, followed by a map $\tilde{n}(i)$ that associates to each basis $i$ the most probable outcome of Alice for that basis.
{Note that this also means that the maximum guessing probability in the classical scenario does not depend on the labelling of the outcomes, since the labelling does not change the probability of the most probable outcome.} Formally:
\begin{align*}
     \Tilde{n}(i)&:=\argmax_{j\in\{0,\ldots,d-1\}} P_A(j|U_i),\\
    P_A(j|U_i)&=\Tr \left(\rho_B U_i \ketbra{j}{j} U_i^\dag\right),\\
    M_i&= \sum\limits_{\substack{i=0  :\Tilde{n}(i)=j}}^{d-1}  \ketbra{i}{i}.\\
\end{align*}
With these definitions we can state the problem as follows: 
\begin{align*}
     &P_g^c:=\\
     &=\max_{\rho_B,\{M_k\}_{k=0}^{d-1}}\sum_{k=0}^{d-1}  \!\Tr_{AB}\!\left[\!\left(\!\rho_B\!\otimes\!\frac{\mathbb{1}}{d} \right)\!CU\!\left(\ketbra{k}{k}\!\otimes\! M_k\right)CU^{\dag}\right]
     \\
     &=\max_{\rho_B} \max_{n_0,n_1,\ldots,n_d} \frac{1}{d}\Tr \left(\sum_{j=0}^{d-1} \rho_B U_j \ketbra{n_j}{n_j} U_j^\dag\right)
     \\
    &=\frac{1}{d} \max_{\rho_B} \Tr\left( \sum_{j=0}^{d-1} \rho_B U_j \ketbra{\Tilde{n}(j)}{\Tilde{n}(j)} U_j^\dag\right)
    \\
    &=\frac{1}{d} \lambda_\text{max} \left[\sum_{j=0}^{d-1} U_j \ketbra{\Tilde{n}(j)}{\Tilde{n}(j)} U_j^\dag\right].\\
    \end{align*}
    where $\lambda_\text{max}[T]$ is the largest eigenvalue of a matrix $T$.
    For small dimensions, the maximum probability can be found by evaluating all possible mappings $\Tilde{n}(j)$ (there are $d^d$ of them). This however quickly becomes infeasible, therefore we look for an upper boound:
    \begin{align*}    
    P_g^c=\frac{1}{d}\left(1+ \lambda_\text{max} \left[\sum_{j=0}^{d-1} U_j \ketbra{\Tilde{n}(j)}{\Tilde{n}(j)} U_j^\dag -\mathbb{1} \right]\right);
\end{align*}
 to simplify the notation we define 
\begin{align*}
    T_j&:= U_j \ketbra{\Tilde{n}(j)}{\Tilde{n}(j)} U_j^\dag -\frac{\mathbb{1}}{d},\\
    T&:=\sum_{j=0}^{d-1} T_j,
\end{align*}
   which satisfy the following properties: 
\begin{align*}
    \Tr (T_j) &=0 \hspace{0.3cm} \forall j\in\{0,\ldots,d-1\},\\
\Tr (T_i^\dag T_j) &=0 \hspace{0.3cm} \forall i\neq j\in\{0,\ldots,d-1\},\\
\Tr (T_j^2)&= \frac{d-1}{d} \hspace{0.3cm} \forall j\in\{0,\ldots,d-1\},\\
\Tr (T^2) &= \Tr \left(\sum_{i,j=0}^{d-1} T_i^\dag T_j\right) \\&= \sum_{i=0}^{d-1} \Tr \left(T_i^\dag T_i\right) +\sum\limits_{\substack{{i,j=0}\\i\neq j}}^{d-1} \Tr  \left(T_i^\dag T_j\right) \\&=d-1.\\
\end{align*}
The guessing probability with a classical coin can be then expressed as
\begin{align*}
 P_g^c:&=\frac{1}{d}\left(1+ \lambda_\text{max} \left[ T \right]\right).
\end{align*}
Since $T$ is trace-less and Hermitian, its largest eigenvalue is positive. We then use the following inequality:
\begin{align*}
    \Tr (T^2) &= \lambda_\text{max}^2 \left[ T \right] \Tr \left(\frac{T^2}{\lambda_\text{max}^2 \left[ T \right]}\right)\\
    &\geq \lambda_\text{max}^2 \left[ T \right]  \left(1+ \min_{S\in M_{d-1}:\Tr S = -1 } \Tr (S^2)\right)\\
    &=\lambda_\text{max}^2  \left[ T \right]\left(1+ \frac{1}{d-1}\right)\\
    &=\lambda_\text{max}^2 \left[ T \right] \frac{d}{d-1};
\end{align*}
where we denoted the space of Hermitian matrices of order $d-1$ by $M_{d-1}$. Substituting the trace of $T^2$ we get the desired upper bound:
\begin{align*}
\lambda_\text{max}  \left[ T \right] &\leq \frac{d-1}{\sqrt{d}},\\
P_g^c &\leq \frac{1}{d}\left(1 +  \frac{d-1}{\sqrt{d}}\right).
\end{align*}

\section{Numeric search}\label{app:numSearch}

\subsection{Classical coin}
When considering a classical coin, the optimal strategy is given by searching over all possible maps $\tilde{n}~:~\mathbb{Z}_d \rightarrow \mathbb{Z}_d $ and taking the largest eigenvalue of the matrix $T=\sum_{j=0}^{d-1} U_j \ketbra{\Tilde{n}(j)}{\Tilde{n}(j)} U_j^\dag -{\mathbb{1}}$. There are $d^d$ such mappings, and we could perform this extensive search for $d=2,3,5,7$, obtaining exact bounds for these dimensions. In other dimensions lower bounds were obtained by applying the see-saw algorithm (\ref{algorithm}) with $\rho_C=\mathbb{1}/d$ and randomized initial points. This algorithm tends to get stuck in local maxima; however in the dimensions in which we could perform the extensive search we observed that the see-saw algorithm returned the maximum value more often then by a random sampling of $\tilde{n}$ in the space of maps $\mathbb{Z}_d \rightarrow \mathbb{Z}_d $.
\subsection{Quantum coin}
For the quantum coin, for the convergence parameter $\eps$ small enough ($10^{-6}$) we didn't observe convergences to local maxima different from the global maximum.\\
Differently from the classical case, the choice of unitaries changes the value of the maximum. We then search for the smallest such value among all possible unitary constructions. The space over which we search is given by choosing $d+1$ unitaries out of the $d+1$ available from the WF construction, and by relabeling, i.e. applying a permutation matrix to each unitary. For $d=3,5$ we searched over all possible permutations, for $d=7$ we only considered cyclic permutations, while for higher dimension we randomly sampled over the space of permutation matrices. Each search is performed for all $d+1$ choices of unitaries. We observed that the WF unitaries give the lowest value when the excluded unitary is the identity.

\bibliography{biblioGG}{}
\end{document}